\documentclass[conference]{IEEEtran}
\IEEEoverridecommandlockouts
\usepackage{cite}
\usepackage{amsmath,amssymb,amsfonts}
\usepackage{algorithmic}
\usepackage{graphicx}
\usepackage{textcomp}
\usepackage{xcolor}
\usepackage{tabularx}

\usepackage{listings}
\usepackage{color}

\definecolor{dkgreen}{rgb}{0,0.6,0}
\definecolor{gray}{rgb}{0.5,0.5,0.5}
\definecolor{mauve}{rgb}{0.58,0,0.82}

\def\BibTeX{{\rm B\kern-.05em{\sc i\kern-.025em b}\kern-.08em
    T\kern-.1667em\lower.7ex\hbox{E}\kern-.125emX}}
\begin{document}

\title{VuLASTE: Long Sequence Model with Abstract Syntax Tree Embedding for vulnerability Detection\\
}


\author{
    \IEEEauthorblockN{
    Botong Zhu, Huobin Tan \IEEEauthorrefmark{1}
    \thanks{\IEEEauthorrefmark{1} Huobin Tan is the corresponding author.}
    }
    \IEEEauthorblockA{
    \textit{School of Software} \\
    \textit{Beihang University} \\
      Beijing, China \\
      \{zbtse, thbin\}@buaa.edu.cn 
    }
} 

\maketitle

\begin{abstract}
In this paper, we build a model named VuLASTE, which regards vulnerability detection as a special text classification task. To solve the vocabulary explosion problem, VuLASTE uses a byte level BPE algorithm from natural language processing. In VuLASTE, a new AST path embedding is added to represent source code nesting information. We also use a combination of global and dilated window attention from Longformer to extract long sequence semantic from source code. To solve the data imbalance problem, which is a common problem in vulnerability detection datasets, focal loss is used as loss function to make model focus on poorly classified cases during training.
To test our model performance on real-world source code, we build a cross-language and multi-repository vulnerability dataset from Github Security Advisory Database. On this dataset, VuLASTE achieved top 50, top 100, top 200, top 500 hits of 29, 51, 86, 228, which are higher than state-of-art researches.
\end{abstract}


\section{Introduction}


Vulnerability detection is the process of identifying and locating vulnerabilities, which are weaknesses or security flaws in software that can be exploited by attackers to gain unauthorized access or perform malicious actions. Automatic vulnerability detection in software is the process of using tools and techniques to automatically identify and locate vulnerabilities in software systems.


The naturalness hypothesis of software provides a theoretical basis for applying natural language models to source code downstream tasks. This suggests natural language models may be an effective alternative to traditional vulnerability detection methods, such as static detection.

Hindle et al. \cite{hindle2016naturalness} provided empirical evidence of programming language naturalness, and their research suggests that programming languages are even more regular than natural languages. Basing on Hindle et al.'s \cite{hindle2016naturalness} result, Allamanis et al. \cite{microsoft_survey} proposed the naturalness hypothesis. 

The naturalness hypothesis of software suggests that software is a form of human communication and that software systems have similar statistical properties to natural language corpora. This aspect of the hypothesis proposes that software systems can be regraded as a type of human language.

The naturalness hypothesis suggests that software corpora have similar statistical properties to natural language corpora. For example, just as natural language text exhibits certain patterns of word usage, software systems may exhibit certain patterns of code usage. By analyzing these patterns, deep learning models, especially attention models \cite{vulsniper}, can be created to learn the structure and organization of software systems. These model can be useful in multiple software engineering areas, including vulnerability detection.

With natural language models, vulnerability detection in source code can be regraded as a text classification task \cite{pang2015predicting} \cite{pang2017predicting}. But existing researches have two main limitations \cite{li2019comparative}.

Firstly, most existing researches focus on C/C++, but currently other languages are becoming increasingly popular, so a cross language dataset may be needed. By training models on a diverse set of code samples from multiple languages, models are more likely to learn vulnerability patterns in a wide range of real-world applications, rather than just those that happen to occur in a specific language. And also, a considerable amount of companies and organizations use multiple languages in their software development process \cite{burow1996mixed} \cite{kochhar2016large}, so having a model that can detect vulnerabilities in multiple languages can be more efficient and cost-effective.

Another limitation in existing researches is the lack of large, annotated datasets for training and evaluating models on programming language tasks. While there are a number of open-source code repositories and other sources of code available, it can be difficult to obtain labeled data for tasks such as vulnerability detection. This can make it difficult to train and evaluate models on these tasks, and may limit the performance of these models.

To solve these problems in existing researches, we proposed VuLASTE (\textbf{Vu}lnerability detection \textbf{L}ong sequence model with \textbf{A}bstract \textbf{S}yntax \textbf{T}ree \textbf{E}mbedding), a deep-learning model for vulnerability detection.

The contributions in this study include:

\begin{itemize}
  \item \textit{Building a cross-language vulnerability dataset from open-source software:} This dataset is more representative of real software projects and encompasses multiple languages and software domains.
  \item \textit{Designing an AST path embedding for the transformer embedding layer:} This enables the embedding to handle nesting information in programming language text, which is important for capturing the structural information of the code.
  \item \textit{Adjusting the long sequence attention mechanism to meet the characteristics of programming languages:} long sequence mechanism allows the model to effectively handle longer input sequences, which is important when dealing with code written in different languages.
  \item \textit{Introducing focal loss to solve the data imbalance problem:} We introduce focal loss from the image recognition field to vulnerability detection. Focal loss is a technique that is used to address class imbalance, which is a common issue in vulnerability detection datasets.
\end{itemize}

The structure of this paper is organized as follows. Section II discusses the related work about vulnerability detection with natural language models. Section III explains the detailed design of VuLASTE. Section IV shows the extraction and generation of training and testing data. Section V shows the experiment results of vulnerability detection and ablation. In section VI, the conclusion and summarization are offered.

\section{Related Work}

There have been a number of researches in recent years that have used deep learning techniques to improve the vulnerability detection. 

VulDeePecker \cite{vuldeepecker} is a deep learning based vulnerability detection model for software source code. It uses a combination of traditional static analysis techniques and deep learning methods to identify vulnerabilities in the code.

The process of VulDeePecker starts with splitting the code text into smaller pieces, and this procedure is called programming slicing. Programming slicing is a traditional static analysis technique that involves identifying and extracting specific parts of a program that are relevant to a given goal or concern.

Once the code is split into smaller pieces, VulDeePecker uses word2vec \cite{word2vec} to create embeddings for the code snippets. Word2vec is a deep learning model that is commonly used to generate embeddings for words in natural language text. In VulDeePecker, word2vec is used to create embeddings for code snippets, which can then be used as input to the next step.

The next step in the process is to use a type of recurrent neural network called Bi-LSTM (bidirectional long short-term memory) \cite{blstm} to classify the code snippets as either vulnerable or non-vulnerable. Bi-LSTM is a type of neural network that can handle text before and after certain token. By using a Bi-LSTM to classify the code snippets, VulDeePecker is able to learn and identify patterns in the code that are indicative of vulnerabilities.

SySeVR \cite{sysevr} is a research based on VulDeePecker. SySeVR uses almost the identical data preprocessing to VulDeePecker. In SySeVR, the input is split into Sy and Se pieces. Sy and Se are short for "syntactic" and "semantic" respectively, and are used to refer to the two types of features that the SySeVR model is designed to analyze.



SySeVR supports more machine learning and deep learning models such as MLP and BiGRU (Bidirectional GRU) \cite{gru}. SySeVR has been shown to achieve strong performance on vulnerability detection task, making it a promising approach to this problem.

CodeBERT \cite{codebert} is a transformer-based language model developed by Microsoft Research that is specifically designed to process programming language text. It was trained on a large dataset of source code and natural language documentation, with similar settings to that of multilingual BERT \cite{multibert}. Like BERT, CodeBERT uses a transformer architecture and learns contextual representations of the input text by attending to the relationships between the tokens in the input sequence.

One key feature of CodeBERT is its ability to handle the specific syntactic and semantic features of programming languages, such as variables, functions, and control structures. This makes it well-suited for tasks that involve processing programming language text, including vulnerability detection.

\section{Design of Model}
\begin{figure*}
    \centering
    \includegraphics[width=\textwidth]{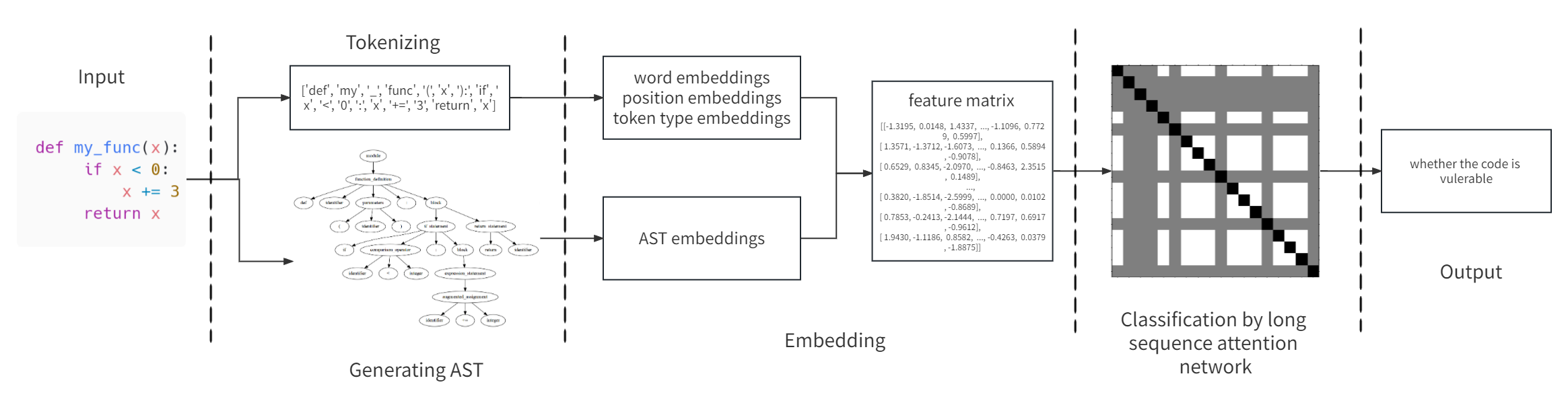}
    \caption{The general framework of VuLASTE}
    \label{fig_framework}
\end{figure*}

The general framework of VuLASTE is shown in figure \ref{fig_framework}. VuLASTE receives source code in function level as input, then tokenizes the text into words and generates AST from text. Based on the tokens, word embeddings, position embeddings, and token type embeddings are generated, and these embeddings are similar with embedding layer of other transformer models. AST path embeddings are generated from the AST, which are used to represent the structural information of the code. By adding these embeddings together, the feature matrix is obtained. This matrix is then input into a long sequence attention model, which is used to classify the code into vulnerable or non-vulnerable cases. In training process, focal loss is introduced to solve the class imbalance problem.

\subsection{Tokenizing}

One difference between tokenizing programming language and natural language text is that there are a large number of identifiers in source code. In programming language text, identifiers are special words that are used to name variables, functions, classes, etc. And if these identifiers are added directly to the vocabulary list, it will cause the vocabulary size to explode.

This is because in natural language text, vocabulary lists usually contain only common words (e.g., nouns, verbs, adjectives, etc.). In programming language text, identifiers may occur in a wide variety of vocabularies with relatively low frequency, so adding all these identifiers to the vocabulary list will increase the size of the vocabulary list dramatically, which may lead to vocabulary list explosion and OOV (Out-Of-Vocabulary) problems \cite{bazzi2002modelling}.

Existing studies such as VulDeePecker, SySeVR, have renamed identifiers as a solution. When tokenizing identifiers including variable names, function names, they are simply renamed to var1, func1, etc. However, the problem of this solution is that if the developer-defined words in identifier are erased, it may lead to the loss of some information of text semantics. Therefore splitting the identifier by word frequency may be a better solution.

In our study, we used byte-level Byte-Pair-Encoding algorithm \cite{bpe} to split identifiers into more common words.

Byte-Pair-Encoding (BPE) is a string partitioning algorithm that can be used to common pair frequently occurring string fragments in a string. Because it captures the phrase and syntactic structure of the language, BPE algorithm is often used for tokenization of language models. The BPE algorithm starts by building a frequency table of all the bytes in the input text, and then it starts merging the most frequent pair of bytes into a new word. This process is repeated until a certain stopping criterion is reached or all string fragments have been merged. In VuLASTE, the BPE algorithm is used to split the input text into subword units, which can then be used as input to the embedding layer.

\subsection{AST generation}

An Abstract Syntax Tree (AST) is a tree representation of the source code of a program. It is an abstract representation of the code that captures the hierarchical structure of the code but abstracts away from some of the details such as the specific characters used. 

In our work, tree-sitter \cite{tree-sitter} is used to generate cross-language ASTs in unified format. Tree-sitter is a parser generator that can be used to create an Abstract Syntax Tree (AST) from source code. This is because Tree-sitter uses a grammar-based approach to parsing, where a custom grammar is defined for each language. This grammar defines the syntax and structure of the language, and is used to generate a parser for that language. By using the unified grammar format for multiple languages, Tree-sitter can generate ASTs in the same format for those languages, allowing for easier comparison and analysis of the code across different languages.

Once the AST is generated, it can be used in ast path embedding to provide structural information about programming language text.

\subsection{Embedding}

\begin{figure}
    \centering
    \begin{lstlisting}
        def my_func(x):
            if x < 0:
              (*@ \fcolorbox{red}{white}{x} @*)+= 3
            return x
    \end{lstlisting}
    \includegraphics[width=\columnwidth]{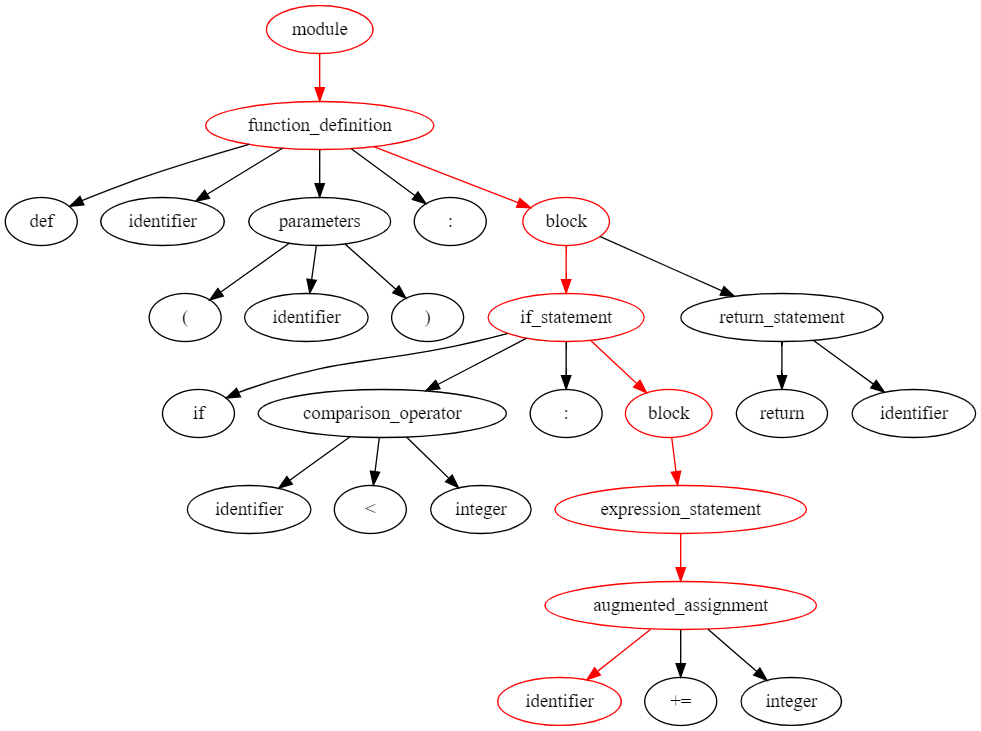}
    \caption{Ast path of example code}
    \label{fig_ast_path}
\end{figure}

Traditional natural language models do not focus on the multi-layered nested structure of programming language when embedding. Compared with natural language text, programming language text is more structured and the structural information has a deeper influence on the semantics.

For example, the same statement in different levels of code blocks may represent different semantics. Based on this characteristic of programming language, this paper adds ast path embedding to the model. The basic concept of ast path embedding is to provide a lightweight representation of ast structure to encode the nested structural information of token in programming language text.

The mathematical representation of ast path embedding is:

Define $AST = (N, E, \phi)$ as the abstract syntax tree of given programming language piece. Define $e_{n_i, n_j}$ as the edge between node $n_i$ and $n_j$.

$$\phi(e_{n_i, n_j}) = <n_i, n_j>$$

Define the path from node $a$ to node $z$ as $W_{a, z}$, 
$$W_{a, z} = <e_{a,b}, e_{b,c}, ..., e_{y,z}>$$

Define $vec(n)$ as the vectorized presentation of node $n$.

For token $t$ corresponding to leaf node $l$, the AST embedding of t is:
$$ AE(t) = \sum\limits_{\substack{e \in W_{root, l} \\ n_i \in \phi(e)}} vec(n_i) $$

And the whole embedding of token $t$ is the sum of AST path embedding, transformer word embedding, position embedding, and token type embedding:

$$ Embedding(t) = WE(t) + PE(t) + TT(t) + AE(t) $$

An example is shown in fig \ref{fig_ast_path}. For the red boxed token y, the ast path embedding is the sum of vectors of red nodes.

\subsection{Long Sequence Attention for vulnerability Detection }
In source code, variables and functions defined previously may have an influence on the semantics of the code that appears later in the sequence, so capturing long dependencies is necessary to accurately understand the meaning of the code. Because of this, a model that only have relatively short memory, such as BiLSTM, may not be sufficient for understanding the code.

Therefore, Long sequence attention is needed to capture long dependencies in source code. It allows the model to attend to information that is farther away in the input sequence. This can be particularly useful in vulnerability detection, where the model needs to understand the relationships between different parts of the code in order to identify patterns that are indicative of vulnerabilities.

The Longformer \cite{longformer} structure is used as our long sequence attention model, and some adjustments are made to meet the specific characteristics of programming language.
Longformer enables the model to attend to a fixed-size window of tokens around the token being processed, which allows the model to attend to longer text under limited memory and computational costs.

\begin{figure}
    \centering
    \includegraphics[width=\columnwidth]{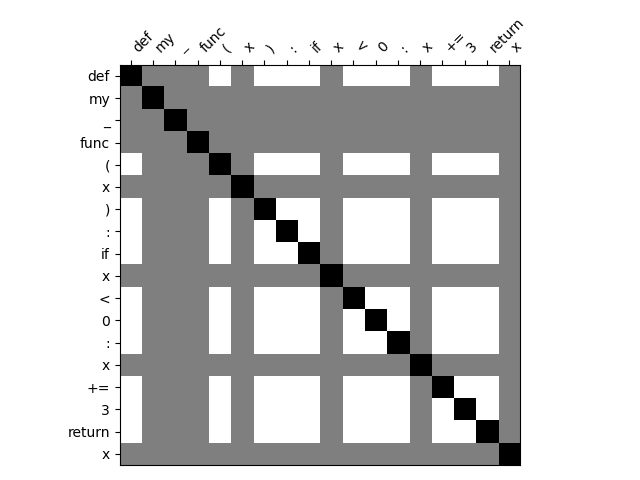}
    \caption{Example of how long sequence attention mechanism may work on programming language}
    \label{fig_attention}
\end{figure}

Longformer uses a combination of global attention and dilated window attention as the replacement of self-attention. This attention mechanism can fit well with the nesting structure of programming language.

One way in which dilated window attention could be used in processing programming language text is by defining a machine learning model that takes as input a sequence of tokens representing the code or natural language documentation of a software system, and produces a prediction or output based on this input. It is similar to sliding window attention in that it operates over a fixed-size window of tokens, but it allows the model to attend to a wider range of tokens by skipping over some tokens within the window.

For example, in the context of vulnerability detection, the global attention may correspond to words that can influence the semantic of entire code file, such as global variables and functions, and local attention may correspond to words that only influence semantic in certain code blocks. An example is shown in figure \ref{fig_attention}. Then the model could use this attention mechanism to attend to security-related keywords or API calls within the code, and use this information to predict whether the system is vulnerable or non-vulnerable.


Usually when processing natural language, Longformer uses a sliding window of 128. Iyer et al. \cite{code-token-len} collected source code pieces from StackOverflow \cite{stackoverflow}, and their study shows the average token length of code blocks varies from 38 to 46, depending on the programming language. So we adjusted the size of sliding window from 128 to 64, which is the smallest power of 2 integer larger than the average length.

\subsection{Class Imbalance}

In the context of vulnerability detection, data imbalance can be a common problem, as vulnerabilities may be rare compared to non-vulnerabilities in a given dataset \cite{8769937}. This can make it difficult for a machine learning model to accurately recognize vulnerabilities, as it may be overwhelmed by the large number of non-vulnerabilities.

Commonly used methods, such as oversampling, can lead to overfitting and also it can be computationally expensive. Therefore, we introduced focal loss \cite{focalloss} from image recognition. Focal loss down-weights the loss for well-classified cases and places more emphasis on the loss for poorly classified cases. This helps the model to focus more on the difficult cases and improve its performance on these cases.

$$ FL(p_t) = - \alpha_t (1-p_t)^\gamma log(p_t) $$

As this formula shows, focal loss may be a better solution to this problem because it allows the model to focus more on the vulnerabilities and less on the non-vulnerabilities, which can improve the model's performance on the vulnerabilities. This can be especially important in situations where the cost of missing a vulnerability is high, as it can help the model to identify as many vulnerabilities as possible, even if it means that it may also produce some false positives.

\section{Data}
The two main options for existing vulnerability detection datasets are manually generated datasets, and data from actual projects. Both options have their own disadvantages, therefore a new cross-language vulnerability dataset may be needed.

Manually generated datasets, such as Juliet \cite{juliet} (used by Russell et al.) and SARD \cite{sard} (used as part of dataset by VulDeePecker \cite{vuldeepecker} and SySeVR \cite{sysevr}), have the advantage of having the equal number of positive and negative examples, which can help to avoid data imbalance.

However, Chakraborty et al.'s research \cite{reveal} shows that \cite{juliet} and SARD \cite{sard} differ significantly from actual software projects.
 This can cause models trained on these datasets to learn features that may not be applicable to actual software projects.

On the other hand, actual software projects can provide a more realistic representation of vulnerabilities in development.
The NVD database \cite{nvd} is a widely used vulnerability database of real-world software. Each vulnerability reported has a CVE serial in NVD database, making the records easy to classify and query. 


Previous studies, such as VulDeePecker \cite{vuldeepecker} and SySeVR \cite{sysevr}, have chosen to identify vulnerable code by scanning GitHub commit messages containing string "cve". This method can be a way to find commits that may be related to code vulnerabilities and extract the vulnerability code from the corresponding commits. However, the disadvantage of this method is that git commit messages are often confusingly and inconsistently formatted between projects. And also, due to the slicing tool VulDeePecker and SySeVR used, the dataset they generated only contains C/C++ language.

\subsection{Dataset Construction}

We use GitHub Advisory Database(GHSA) \cite{ghsa} as data source. Items in GHSA are from open-source repositories hosted on GitHub. Developers who fix vulnerabilities in their projects have the option to publicly disclose their vulnerability information and fix patches to the GitHub community. Since the majority of current well-known open-source projects are hosted on Github, the vulnerabilities included in the GHSA database have a relatively high consistency with the NVD database entries involving open-source software \cite{buhlmann2022developers} \cite{horawalavithana2019mentions}.



The following procedures are used in generating the code vulnerability dataset.
And for each case, the codes are extracted at function level.

\begin{enumerate}
    \item Determine the target programming language need to be extracted, the programming languages can be various.
	\item Download the full list of identified vulnerabilities from the GHSA database, and exclude records that do not have a fix patch or patch is not available. 
	\item For each git commit patch, filters the files involved in the git commit to exclude non-code files or non-target language files.
	\item For each code file modified by a git commit, extracts the unfixed version and the fixed version of the code separately.
	\item Generates an AST (abstract syntax tree) of the unfixed and the fixed versions of each code file separately, and slices the code files at the function level according to the structure of the AST, with each function as one piece of data in the dataset. 
	\item The unfixed version is labeled as vulnerable, while the fixed version is labeled as non-vulnerable. Some unmodified functions are also present in both fixed and unfixed versions, these functions are labeled as non-vulnerable.
\end{enumerate}



The dataset in this paper is generated in 2022 March. The finally generated dataset contains a variety of popular programming languages (C, C++, Java, Python, Go), and the total size of dataset is 11072.

The code length statistics (with the number of characters as data length) are shown in table \ref{tab_length_statistics}. As can be seen, the dataset contains not only short code pieces, but also a large amount of long code text, which is consistent with what is found in software development.

In the extraction process, some unchanged code is also included in the git diff messages, making size of the final extracted negative cases (non-vulnerable codes) larger than positive cases (vulnerable codes). As the result, negative cases are 8601, positive cases are 2471.

\begin{table}[]
    \begin{center}
        \begin{tabular}{lcc}
        \hline
        Length & & count \\
        \hline
        {[}0.0, 512.0) &  & 6200 \\
        {[}512.0, 1024.0) &  & 1616 \\
        {[}1024.0, 2048.0) &  & 1560 \\
        {[}2048.0, 5096.0) &  & 1180 \\
        {[}5096.0, inf) &  & 516 \\
        \hline
        \end{tabular}
    \end{center}
    \caption{Length statistics of dataset}
    \label{tab_length_statistics}
\end{table}



\section{Experiments}
\subsection{Experiment Settings}

\begin{figure*}
    \centering
    \includegraphics[width=\textwidth]{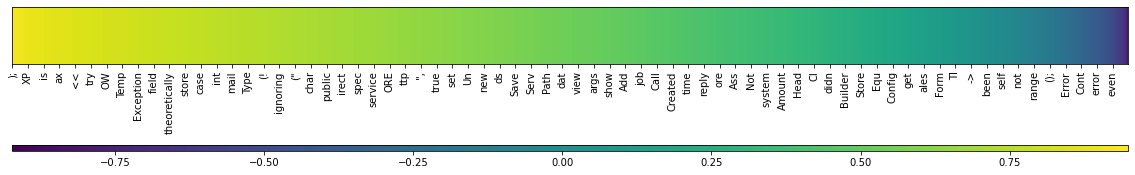}
    \caption{Heat visualization of attention weights}
    \label{fig_heatmap}
\end{figure*}

\subsubsection{Dataset}

For all experiments, we use the dataset generated in the Data section. Same with common approach, we randomly split the dataset into train, validation and test dataset, with corresponding size ratio of 60\%, 20\%, 20\%.

\subsubsection{Environment}

We run our experiments on a Ubuntu Server 22.04 machine with i9-9820X CPU, 128G RAM, dual Nvidia RTX 3090 GPU. The Python version is 3.8.

\subsection{Vulnerability Detection Experiments}

\subsubsection{Baseline}

We chose the following models as baseline. 

SySeVR-BGRU is the best performance model in SySeVR paper \cite{sysevr}. VulDeePecker \cite{vuldeepecker} has the same model architecture with SySeVR-BLSTM. SySeVR takes into account both syntactic and semantic features of the code, which can provide a more comprehensive view of the system and improve the model's performance, making it a good benchmark for comparison.

CodeBERT \cite{codebert} is a bimodal pre-trained model that has been specifically designed for multiple downstream code-related tasks, so it may be more effective at handling the characteristics of code (such as its syntax and vocabulary) than more general-purpose models. CodeBERT has been trained on a large and diverse dataset of code, so it may be able to capture the complex patterns and relationships present in code more effectively than models trained on smaller or more homogeneous datasets.

VUDDY \cite{vuddy} is a representative research of traditional static detection. VUDDY declares that by using clone detection to find codes similar to previously found vulnerable codes, vulnerable in new software projects can be identified. The clone detection method of VUDDY is to calculate the AST similarity of target code and codes in vulnerability database, and target code is considered vulnerable if similarity is over threshold.

\subsubsection{Metrics}

The purpose of the vulnerability detection model is to select potentially vulnerable code from software projects that need to be reviewed by human experts. Therefore we use top-k hits as the main metric of evaluating the performance of models, and recall as the second evaluation method.

One reason to use top-k hits as a metric is that it can be more forgiving of false negatives. In a vulnerability detection task, it is often more important to identify as many vulnerabilities as possible, even if this means that the model may also produce some false positives.

\subsubsection{Experiment Results}

The experiment results are shown in table \ref{tab_exp1}. As the table shows, our model performs better than state-of-art researches.

Since SySeVR is derived from VulDeePecker, SySeVR-BLSTM has the same architecture with VulDeePecker. And SySeVR-BGRU is an improved model basing on VulDeePecker. The results of these two models may suggest attention mechanism can replace the difficult code slicing process in SySeVR/VulDeePecker.

VUDDY classified all test cases as non-vulnerable, which may indicate VUDDY meets challenge in detecting real-world vulnerabilities.

To understand how the model is weighting different parts of the input code when making a prediction, attention weights are visualized in fig \ref{fig_heatmap}. As the picture shows, the model mainly focus on high-risk operations that may cause vulnerabilities, such as left shift and Exception handling. This result may suggest that the VuLASTE is able to identify these patterns effectively and correctly. This is a great indication that the model is working effectively and has learned to recognize the characteristics of vulnerable code.

\begin{table}[]
    \begin{center}
        \resizebox{\columnwidth}{!}{
        \begin{tabular}{ccccccc}
        \hline
        Model & hits@50 & @100 & @200 & @500 & recall & f1 \\
        \hline
        \textbf{VuLaste} & \textbf{29} & \textbf{51} & \textbf{86} & \textbf{228} & 0.4820 & \textbf{0.4801} \\
        VulDeePecker & 6 & 12 & 33 & 77 & \textbf{0.5404} & 0.1941 \\
        SySeVR-BGRU & 5 & 15 &37  & 85 & 0.3826 & 0.2410 \\
        CodeBERT & 3 & 6 & 24 & 63 & 0.4463 & 0.4203 \\
        VUDDY & 0 & 0 & 0 & 0 & - & 0.0000 \\
        \hline
        \end{tabular}}
    \end{center}
    \caption{The metrics of different models.}
    \label{tab_exp1}
\end{table}

\begin{table}[]
    \begin{center}
        \resizebox{\columnwidth}{!}{
        \begin{tabular}{ccccccc}
        \hline
        Model & hits@50 & @100 & @200 & @500 & recall & f1 \\
        \hline
        \textbf{VuLaste} & \textbf{29} & \textbf{51} & \textbf{86} & \textbf{228} & \textbf{0.4820} & \textbf{0.4801} \\
        no AST & 3 & 15 & 27 & 66 & 0.4983 & 0.3582 \\
        self attention & 4 & 15 & 28 & 81 & 0.5369 & 0.3794 \\
        cross entropy & 0 & 0 & 0 & 0 & - & 0.0000 \\
        \hline
        \end{tabular}}
    \end{center}
    \caption{The results of ablation study.}
    \label{tab_ablation}
\end{table}

\subsection{Ablation Study}

The results of ablation study are shown in table \ref{tab_ablation}.

The 'no AST' experiment means model without AST path embedding. This experiment shows that the AST path embeddings are important for capturing the structural information of the code, which is useful for identifying patterns or anomalies that may indicate vulnerabilities.

The 'self attention' experiment replace the long sequence attention mechanism with traditional transformer self attention. This result shows that the long sequence attention mechanism is effective at handling longer input sequences, which is important when dealing with identifiers in programming language texts.

The 'cross entropy' experiment removed focal loss from the model, and used cross entropy instead. The result of removing focal loss suggest the model may struggle to learn vulnerable patterns if not deal with data imbalance problem.

In summary, the results of the ablation experiments demonstrate that each part of the model is important and contribute to the overall performance. The AST path embeddings, long sequence attention mechanism and focal loss are all important for the model to effectively identify vulnerabilities in the code.

\section{Conclusion}

In this paper, we proposed VuLASTE, a deep learning model to detect vulnerable codes. To deal with vocabulary explosion problem, our model use bpe algorithm from natural language processing in tokenizing. This model also adds AST path embedding to provide a lightweight representation for programming language nesting structure. To replace the program slicing method, we use a long sequence attention mechanism from Longformer, combining global attention and windowed attention, to capture long-term semantic in source code. We also extracted a dataset from real-world open source repositories from Github Security Advisory Database. Experiment results show that comparing with existing researches, VuLASTE can better select source code pieces that may be vulnerable when candidate number is limited.

\section*{Acknowledgment}

The authors would like to thank Zilie Wang, who provided excellent idea with experiments design and helped maintain our GPU server, Yufeng He, who implemented the software to extract vulnerabilities from GitHub Advisory Database, Tengqing Jiang, who fixed multiple fatal errors in SySeVR data preprocessing code. We would also like to express our gratitude to GitHub community for providing GHSA database that was essential to this work.


The authors would also like to thank the anonymous reviewers for their helpful comments and suggestions, which have greatly improved the quality of the paper.

\bibliographystyle{IEEEtran}
\bibliography{ref}

\end{document}